# Protocol for an Observational Study on the Effects of Early-Life Participation in Collision Sports on Later-Life Cognition in a Sample of Monozygotic and Dizygotic Swedish Twins Reared Together and Twins Reared Apart

Jordan Weiss[1], Amanda R. Rabinowitz[2], Sameer K. Deshpande[3], Raiden B. Hasegawa[4,5],

Dylan S. Small[4*]

## Abstract

**Background:** A large body of work links traumatic brain injury (TBI) in adulthood to the onset of Alzheimer's disease (AD). AD is the chief cause of dementia, leading to reduced cognitive capacity and autonomy and increased mortality risk. More recently, researchers have sought to investigate whether TBI experienced in early-life may influence trajectories of cognitive dysfunction in adulthood. It has been speculated that early-life participation in collision sports— a leading cause of concussions among adolescents—may lead to poor cognitive and mental health outcomes. However, to date, the few studies to investigate this relationship have produced mixed results. We propose to extend this literature by conducting a prospective study on the effects of early-life participation in collision sports on later-life cognitive health using the Swedish Adoption/Twin Study on Aging (SATSA). The SATSA is unique in its sampling of monozygotic and dizygotic twins reared together (respectively MZT, DZT) and twins reared apart (respectively MZA, DZA). **Methods and Analysis:** The proposed analysis is a prospective study of 660 individuals comprised of 270 twin pairs and 120 singletons. Seventy-eight (11.8% individuals reported participation in collision sports. Our primary outcome will be an indicator of cognitive impairment determined by scores on the Mini-Mental State Examination (MMSE). We will also consider several secondary cognitive outcomes including verbal and spatial ability, memory, and processing speed. Our sample will be restricted to individuals with at least one MMSE score out of seven repeated assessments spaced approximately three years apart. We will adjust for age, sex, and education in each of our models.

[1] Population Studies Center and the Leonard Davis Institute of Health Economics, University of Pennsylvania
[2] Moss Rehabilitation Research Institute
[3] Computer Science and Artificial Intelligence Laboratory, Massachusetts Institute of Technology
[4] Department of Statistics, The Wharton School, University of Pennsylvania
[5] The author contributed to this work while at the University of Pennsylvania. He is currently at Google Inc
* Corresponding author. E-mail: **dsmall@wharton.upenn.edu**

**Background**

Sports-related concussions have emerged in the past decade as a leading healthcare issue for children and adolescents in the US and abroad. The Centers for Disease Control and Prevention (CDC) define a concussion as a mild traumatic brain injury (mTBI) resulting from a direct impact or forceful motion to the head (Schuchat, Houry, & Baldwin, 2018). Although concussions can occur in any sport, they are most common in collision sports such as football, hockey, and soccer (Harmon et al., 2013). In the US, it is estimated that children and adolescents account for 78.6% of the sports-related head trauma cases treated in emergency departments (Gaw & Zonfrillo, 2016) although it is believed that nearly 50% of sports-related concussions go unreported (Harmon et al., 2013). In light of this underreporting and the susceptibility of this population to the effects of sports-related concussions (Halstead & Walter, 2010; Patel & Reddy, 2010; Yeates, 2010), there is a growing need to understand whether early-life participation in collision sports has implications for later-life cognitive function.

Our goal in the proposed analysis is to expand on prior studies by investigating the link between early-life participation in collision sports and late-life cognitive impairment. We capitalize on the Swedish Adoption/Twin Study of Aging (SATSA; Finkel & Pedersen, 2004), an ongoing and longitudinal study representing the Swedish population age 50+ (Pedersen et al., 1991) with survey years in 1984, 1987, 1990, 1993, 2004, 2007, and 2010. The SATSA contains data on a subset of same-sex monozygotic and dizygotic twin pairs from the population-based Swedish Twin Registry (STR) including all twin pairs reared apart and a sample of twins reared together matched on gender, date of birth, and county of birth. The SATSA is unique in its twin design, rich retrospective data, and extensive in-person cognitive assessment which includes verbal and spatial ability, memory, and processing speed. In the 1993 survey wave, responders were asked to report whether they were involved in any sport (e.g., football, ice hockey, or boxing) that may involve a hit on a head (hereafter collision sports). Examining twin pairs concordant and discordant for participation in collision sports allows us to control for biological and familial influences while comparing cognitive outcomes.

**Eligibility and Exclusion Criteria**

The Swedish Adoption/Twin Study of Aging (SATSA) is a longitudinal study comprised of all same-sex twin pairs from the population-based Swedish Twin Registry who indicated they were separated from their co-twin prior to age 11 and reared apart (Finkel and Pedersen 2004). The SATSA also includes a control sample of twins reared together matched on age, sex, and county of birth in Sweden. Sample ascertainment and study procedures have been described in detail elsewhere (Finkel and Pedersen 2004; Pedersen 2015). Briefly, SATSA researchers collected information about health, lifestyle, and environment from eligible participants by mail questionnaire when the study began in 1984 (response rate 70.7%). Twin pairs aged 50 years or older were invited to complete in-person tests to assess their health and cognitive abilities during a four hour visit conducted by a trained and registered nurse practitioner. Questionnaires and in-person visits continued approximately every three years with seven follow-up periods (1987,

1990, 1993, 2004, 2007, 2010, 2014) until the study ended in 2014. SATSA was approved by the Ethics Committee at Karolinska Institutet (Finkel and Pedersen 2004; Pedersen 2015). All participants provided informed consent for participation.

Our primary sample (Sample 1) is obtained from a restricted data file with linked mortality records prepared by SATSA investigators for this specific investigation. This sample contains 660 respondents aged 50 years or older with full covariate information who completed at least one in-person interview over the survey period and responded to the 1993 mail-out questionnaire (Questionnaire 4) wherein participation in collision sports was assessed. Sample 2 is prepared using the public-use SATSA data files with the same exclusion criteria applied, resulting in a sample of 662 respondents. Fisher's test revealed no statistically significant differences in the missingness of cognitive scores by participation in collision sports (P value = 1).

**Study Outcomes**
Our primary outcome is cognitive impairment suggestive of dementia or mild cognitive impairment (MCI) as assessed through the MMSE. The MMSE is commonly used to assess cognitive functioning, track changes in cognitive function over time, and screen individuals for cognitive impairment (Folstein, Folstein, & McHugh, 1975). The SATSA implemented the MMSE in its baseline assessment in 1984 and in each wave that followed through 2010. Table 1 shows descriptive statistics for MMSE scores over the survey period for Sample 1. We will dichotomize MMSE scores such that scores between 0 and 27 are coded as impaired (1) and scores between 28 and 30 are coded as not impaired (0). The traditional MMSE cutoff score is 24, but higher cutoff scores have been proposed to increase diagnostic accuracy in individuals with higher levels of education, and at earlier stages of dementia severity. In the absence of a conclusively defined cutoff, we will replicate our semi-parametric illness-death model for MMSE using two additional cutoffs—24 and 29—as recommended in the literature (Bassett & Folstein, 1991; Rajji et al., 2009).

In conjunction with the MMSE, the SATSA administered an extensive cognitive battery which will allow us to assess whether participation in collision sports is associated with level of or change in trajectories of cognitive performance over five domains of cognition, including general cognitive ability, verbal ability, spatial ability, memory, and processing speed. Table 2 shows descriptive statistics for the cognitive domains in our sample of eligible responders in Sample 2.

**Primary Analysis**
We intended to use a three-state illness death model to study the association between participation in collision sports and cognitive impairment while accounting for the competing risk of death. Due to non-convergence of our models, we instead used the Fine-Gray model. Both approaches are able to account for the interval censoring of cognitive impairment between survey waves and the competing risks of death and study drop-out (Fine and Gray 1999; Touraine, Helmer and Joly 2016). Competing risks arise when the occurrence of a particular event (e.g., death) precludes the event of interest (e.g., dementia). In this framework, the subhazard (i.e., the hazard function of the subdistribution) for dementia is used to associate covariates with the cumulative probability of dementia while accounting for the competing risk of mortality, resulting in a cumulative incidence function (CIF) for dementia.

Our primary outcome is a binary indicator of dementia as defined by an MMSE score of 27 or below. Sensitivity analyses will be conducted using the aforementioned MMSE cutoff scores of 24 and 29. Age will be used as the time scale of analysis and all models will be adjusted for sex and education.

**Secondary Analysis**

We will estimate trajectories of cognitive performance across multiple domains (global cognition, verbal ability, spatial ability, memory, processing speed) using a series of growth curve models to examine the association of participation in collision sports with mean cognitive performance and change over time. Growth curve models allow for the exploration of both intra-individual change and individual differences in the nature of that change (Curran, Obeidat and Losardo 2010; Singer and Willet 2003).

We will estimate a first model that includes a fixed effect for collision sports participation to investigate the association between collision sports participation and mean cognitive abilities at baseline (i.e., the intercept). A second model will include a fixed effect for collision sports as well as linear and quadratic age interaction terms to investigate whether and how participation in collision sports may affect cognitive abilities over time (i.e., the slope). All models will be adjusted for age (centered at 65), sex, and education. We will account for the hierarchical structuring of repeated measures nested within the individual nested within twin pairs by modeling the latter component as random effects. All models will be estimated using full-information maximum likelihood. We will cluster the standard errors to account for the twin structure.

**Table 1**. Descriptive statistics for the MMSE scores across the study period.

| Wave[1] | n | Mean | Standard deviation | Min. | Max. |
|---|---|---|---|---|---|
| Wave 1 | 464 | 28.1 | 1.6 | 22 | 30 |
| Wave 2 | 478 | 28.4 | 1.3 | 23 | 30 |
| Wave 3 | 503 | 27.9 | 1.9 | 13 | 30 |
| Wave 5 | 484 | 27.2 | 2.8 | 8 | 30 |
| Wave 6 | 406 | 27.2 | 3.1 | 4 | 30 |
| Wave 7 | 347 | 27.1 | 3.7 | 7 | 30 |
| Wave 8 | 308 | 26.6 | 3.9 | 3 | 30 |
| Wave 9 | 274 | 26.4 | 4.78 | 0 | 30 |

Notes. MMSE, Mini-Mental State Examination. Shown for Sample 1.
[1] Due to reasons related to funding, a telephone interview was used in place of in-person testing for Wave 4 and thus the cognitive assessment was not conducted for all individuals (Pedersen et al., 1991).

**Table 2**. Descriptive statistics for the cognitive domains across the study period.

| Wave | General cognitive ability | Verbal ability | Spatial ability | Memory | Processing speed |
|---|---|---|---|---|---|
| Wave 1 | | | | | |
| n | 466 | 466 | 466 | 466 | 466 |
| Mean (SD) | 51.1 (9.5) | 51.2 (9.6) | 50.8 (9.9) | 50.9 (9.6) | 51.2 (9.1) |
| Wave 2 | | | | | |
| n | 477 | 477 | 477 | 477 | 477 |
| Mean (SD) | 52.3 (9.6) | 52.4 (9) | 52 (9.8) | 52 (9.7) | 51.5 (9.7) |
| Wave 3 | | | | | |
| n | 503 | 503 | 503 | 503 | 503 |
| Mean (SD) | 53.3 (10.2) | 52.8 (9.3) | 51.8 (10.6) | 52.1 (10.6) | 51.7 (10.7) |
| Wave 5 | | | | | |
| n | 484 | 484 | 484 | 484 | 484 |
| Mean (SD) | 52.4 (10.3) | 53.3 (10) | 51.4 (11.1) | 50.8 (9.9) | 48.8 (11) |
| Wave 6 | | | | | |
| n | 406 | 406 | 406 | 406 | 406 |
| Mean (SD) | 52.8 (10.1) | 53.9 (9.4) | 51.1 (11.1) | 51.1 (9.6) | 49.4 (11.4) |
| Wave 7 | | | | | |
| n | 347 | 347 | 347 | 347 | 347 |
| Mean (SD) | 55.1 (10.1) | 55 (9.3) | 54 (11.1) | 53 (10.2) | 50.5 (11.2) |

Notes. SD, standard deviation. Shown for Sample 2.

[1] Due to reasons related to funding, a telephone interview was used in place of in-person testing for Wave 4 and thus the cognitive assessment was not conducted for all individuals (Pedersen et al., 1991).